\documentclass[12pt]{article}

\usepackage{graphicx}
\usepackage{dsfont}
\usepackage{amssymb}
\usepackage{mathrsfs}

\begin{document}
\title{Observation of negative-frequency waves in a water tank: 
A classical analogue to the Hawking effect?}
\author{Germain Rousseaux$^{1,2}$, 
Christian Mathis$^2$,  
Philippe Ma\"{i}ssa$^2$,\\
Thomas G. Philbin$^{3,4}$,
and Ulf Leonhardt$^3$\\
$^1$ACRI, Laboratoire G\'{e}nimar, 260 route du Pin Montard,\\
B.P. 234, 06904 Sophia-Antipolis Cedex, France\\
$^2$Universit\'{e} de Nice-Sophia Antipolis, 
Laboratoire J.-A. Dieudonn\'{e},\\
UMR CNRS-UNSA 6621,
Parc Valrose,
06108 Nice Cedex 02, France\\
$^3$School of Physics and Astronomy, University of St Andrews,\\
North Haugh, St Andrews KY16 9SS, Scotland, UK\\
$^4$Max Planck Research Group of Optics, Information and Photonics,\\
G\"unther-Scharowsky-Str.\ 1, Bau 24, D-91058 Erlangen, Germany
}
\maketitle
\begin{abstract}
The conversion of positive-frequency waves into negative-frequency 
waves at the event horizon is the mechanism at the heart of the Hawking radiation of black holes.
In black-hole analogues, horizons are formed for waves propagating
in a medium against the current 
when and where the flow exceeds the wave velocity.
We report on the first direct observation of negative-frequency 
waves converted from positive-frequency waves in a moving medium.
The measured degree of mode conversion 
is significantly higher than expected from theory.
\\

\end{abstract}

\newpage

\section{Introduction}

The theory of Hawking radiation
of black holes \cite{Hawking} connects 
three separate disciplines of physics --- 
quantum mechanics, general relativity 
and thermodynamics \cite{Bekenstein} --- 
and has been applied to test potential quantum theories 
of gravity \cite{GSW,Rovelli}.
The radiation of astrophysical black holes is too feeble 
to be detectable, 
but laboratory analogues
\cite{Novello,SUbook,Fiber,Volovik}
of the event horizon 
may demonstrate the physics behind Hawking radiation. 
Most candidates of artificial black holes 
rely on quantum fluids 
\cite{Volovik,Unruh,Visser,Garay,Giovanazzi},
but here we report on an experiment with a classical fluid: 
water \cite{SU}.
A horizon is formed when flowing water exceeds the wave velocity. 
We observed a key ingredient of the classical mechanism 
behind Hawking radiation, 
the generation of waves with negative frequencies
\cite{Hawking,Brout,BD}.
However, the measured conversion of positive into negative-frequency
waves is significantly higher than expected from theory \cite{SU} 
for reasons we do not yet understand.

\begin{figure}[h]
\includegraphics[width=10.0cm]{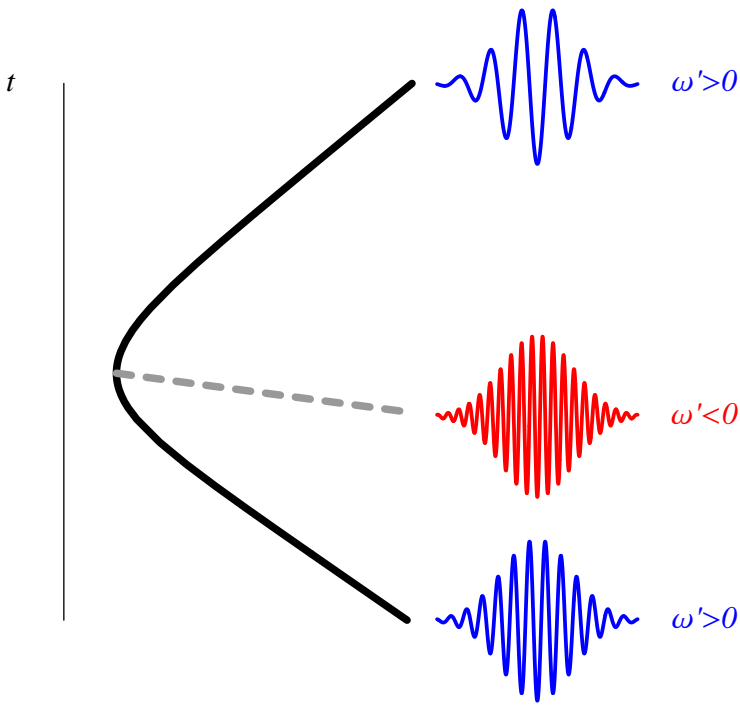}
\caption{{\small
Tracing wave packets backwards in time at the horizon
of a black hole.
Schematic space-time diagram showing
a wave packet escaping into space (top),
potentially reaching an observer.
This wave packet oscillates at positive frequencies,
but it originates from two distinct waves,
one with positive and another one with negative 
frequencies,
shown below the escaping wave packet in
the space-time diagram
(for times in the past).
This mixing of positive and negative frequencies
is the classical root of the quantum Hawking radiation 
\cite{Hawking}. 
Note that the deflection of the incident
waves at the horizon depends on the dispersion properties
of the "space-time medium"
\cite{tHooft,Jacobson,Unruh2,Brout2,CJ}.
In astrophysics, these properties are unknown,
in contrast to laboratory analogues.
}
\label{fig:black}}
\end{figure}

\newpage

\begin{figure}[h]
\includegraphics[width=11.5cm]{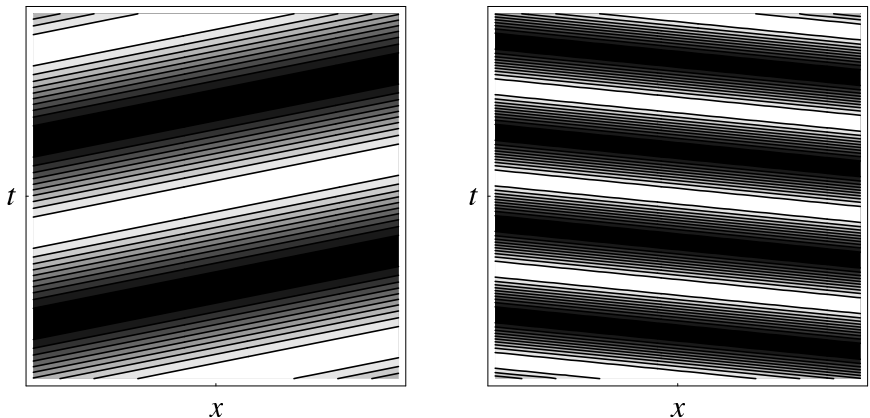}
\caption{{\small
Positive versus
negative-frequency waves.
The left diagram shows the space-time diagram 
of a wave with positive frequency,
while the right diagram shows a negative-frequency wave.
Section 2 explains the physics of  
negative-frequency waves in moving media. 
The pictures show
space-time diagrams of waves in media moving
with uniform speed.
The left diagram displays a wave with positive
wavenumber $k$,
whereas the right diagrams shows
a wave with negative $k$ and 
negative frequency $\omega'$
in the co-moving frame.
}
\label{fig:diagrams}}
\end{figure}

In 1974 Hawking \cite{Hawking} 
predicted that black holes are not black: 
they radiate. The event horizon generates pairs of quanta; 
one particle of each pair emerges into space 
while its partner falls into the singularity. 
The quantum physics of pair creation at horizons
is based on the features of classical wave-packet propagation
\cite{Hawking2,Brout,BD} as follows:
Figure \ref{fig:black} shows a wave packet 
escaping from the horizon. 
In a thought experiment,
Hawking  \cite{Hawking2}
traced such wave packets backwards in time
and realized that they originate from two distinct 
waves: one oscillating with positive frequencies
and another one with negative frequencies.
Note that one can visualize 
negative frequencies in 
the way waves propagate in space and time,
{\it i.e} in space-time diagrams or videos,
but negative frequencies 
do not  directly appear in snapshots of wave packets.
Figure \ref{fig:diagrams}
compares the space-time diagrams of ordinary positive-frequency
waves with the behavior of negative-frequency waves.
The figure shows that the lines of equal phase 
in space-time have negative slopes for
negative frequencies, 
as we discuss in Sec.\ 2. 

\begin{figure}[h]
\includegraphics[width=15.0cm]{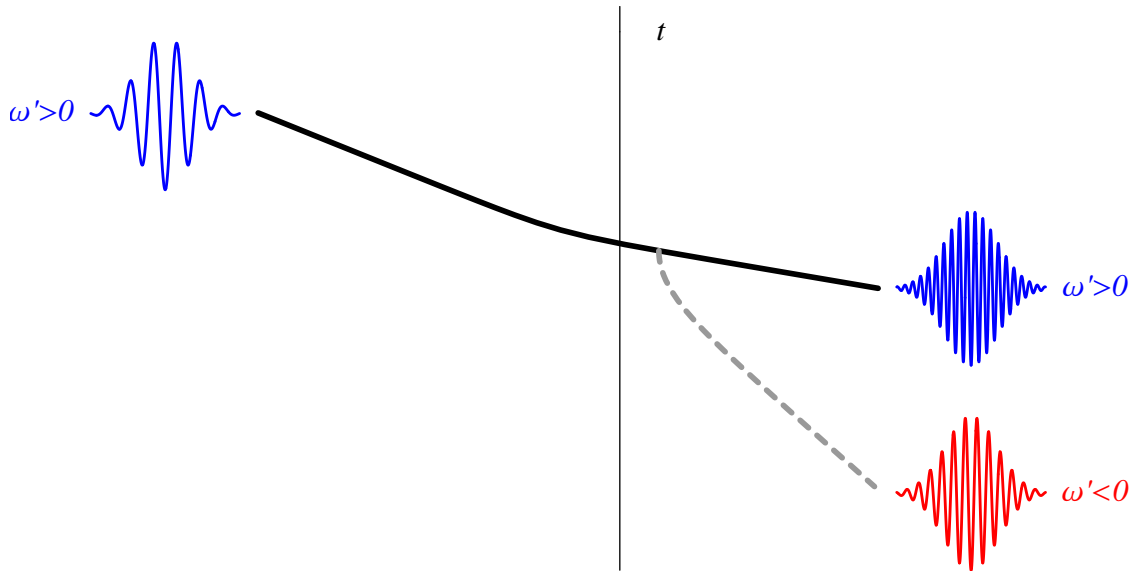}
\caption{{\small
Hawking partner.
Schematic space-time diagram of a wave packet
propagating against the "space-time flow"
on the other side of the horizon,
drifting towards the singularity of the black hole.
Like the wave illustrated in Fig.\ \ref{fig:black}
this wave packet  
originates from waves with positive and negative 
frequencies. 
These waves are mixtures of the escaping waves
of Fig.\ \ref{fig:black}
traced backwards in time;
hence the escaping quanta and the in-falling quanta
form entangled partners.  
}
\label{fig:black2}}
\end{figure}

The distinction between positive and negative frequencies is 
important for quantum fields 
\cite{Hawking2,Brout,BD}: 
the positive frequencies distinguish the annihilation 
and the negative frequencies the creation operators. 
A process that mixes positive and negative frequencies 
thus creates particles; 
the horizon spontaneously emits radiation.
Figure  \ref{fig:black} illustrates the wave packets
of the particles that escape into space;
the particles that fall into the black hole
are shown in figure \ref{fig:black2}.
They originate from mixtures of the 
two initial wave packets of Figure  \ref{fig:black}.
Therefore the created quanta appear in entangled
pairs, one escaping, the other one falling into the singularity. 

Seen from outside, 
the black hole turns out \cite{Hawking2,Brout,BD}
to emit black-body radiation 
with a temperature \cite{Hawking} 
that is proportional to the surface gravity at the horizon, 
or, equivalently, inversely proportional 
to the size of the black hole, the Schwarzschild radius. 
Since Hawking's prediction, 
the radiation of horizons has been regarded as a confirmation 
for black-hole thermodynamics \cite{Bekenstein} 
and as a crucial test case for quantum theories of gravity 
such as superstring theory \cite{GSW} 
and loop quantum gravity \cite{Rovelli}. 

However, near the event horizon, 
fields are subject to frequency shifts beyond the Planck scale
\cite{tHooft,Jacobson,Unruh2,Brout2,CJ},
as Fig.\ \ref{fig:black} schematically illustrates:
the incident wave packets oscillate at significantly
higher frequencies than the outgoing waves.
The mechanism that could limit the frequency shifting 
at the horizon of the astrophysical black hole is unknown.
Hawking radiation may thus depend on as yet unknown physics 
or may not exist at all.  
There is no observational evidence for Hawking radiation 
in astrophysics yet; 
and it seems unlikely that there ever will be 
for practical reasons ---
radiation with characteristic thermal wavelengths 
in the order of the Schwarzschild radius, a few km for solar-mass black holes, 
is obscured by the cm-waves 
of the Cosmic Microwave Background. 

Astrophysical black holes are too large for noticeable 
Hawking radiation, but laboratory analogues 
\cite{Novello,SUbook,Volovik,Fiber}
of black holes offer valuable insights into the mechanism 
of radiating horizons. 
Most analogues are based on a simple idea 
\cite{Volovik,Unruh,Visser}: black holes behave like moving fluids. 
Consider waves with phase velocity $c'$ 
in a medium of flow speed $u$. 
If the magnitude of $u$ exceeds $c'$ 
waves can no longer propagate upstream; 
they are trapped beyond a horizon. 
The horizon creates wave-quanta 
\cite{Novello,SUbook,Fiber,Volovik}, 
the analogue of Hawking radiation \cite{Hawking}, 
with an effective temperature that depends on the flow gradient 
at the horizon, the analogue \cite{Novello,SUbook,Fiber,Volovik} 
of the surface gravity. 
The radiation is only noticeable 
if the temperature of the fluid lies below 
the effective Hawking temperature. 
Superfluids \cite{Volovik} like Helium-3 
or ultracold quantum gases \cite{Garay,Giovanazzi}
may form radiating horizons for their elementary excitations 
and so would moving optical media for photons
\cite{Fiber,LeoReview}.

\begin{figure}[t]
\includegraphics[width=10.0cm]{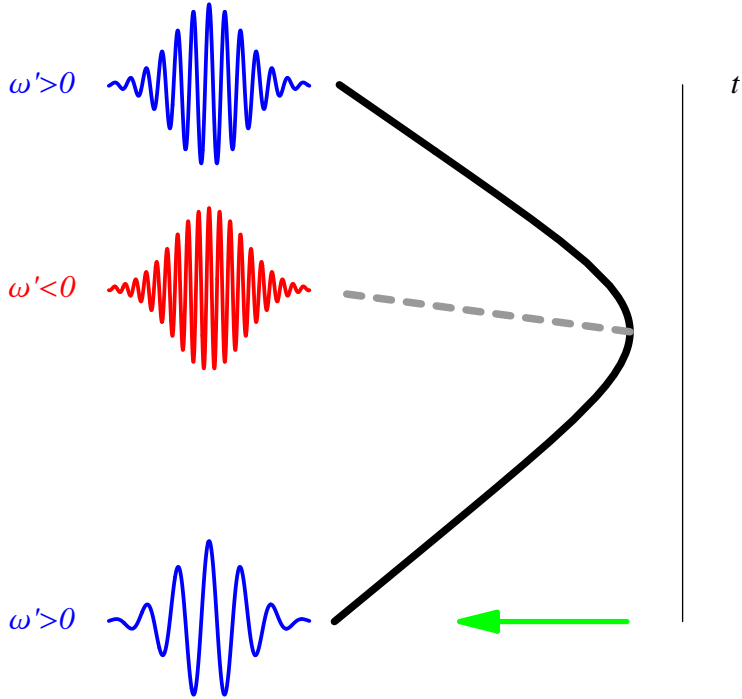}
\caption{{\small
White-hole horizon. 
In order to demonstrate in a laboratory setting 
the tracing of wave packets backwards in time 
at a black-hole horizon, 
one has to time-reverse Fig.\ \ref{fig:black}. 
The time-reversed black hole is the white hole. 
The arrow indicates the direction
of the moving medium 
that establishes a horizon for counter-propagating waves.
}
\label{fig:white}}
\end{figure}

On the other hand, 
at the heart of the Hawking effect 
lies a classical process 
that can be demonstrated with classical fluids such as water: 
the generation of waves with negative frequencies.  
For this, one should reproduce
the characteristic behavior of wave packets at horizons 
traced backwards in time illustrated in Fig.\ \ref{fig:black}.
This is possible with a time-reversed black hole ---
a white-hole horizon --- as shown in Fig. \ref{fig:white}. 
The horizon of the white hole corresponds to the following analogy:
imagine a fast river flowing out into the sea, getting slower.
Waves cannot enter the river beyond the point
where the flow speed exceeds the wave velocity;
beyond this point the river resembles an object 
that nothing can enter, the white hole.
Such wave blocking has been comprehensively 
studied in the fluid-mechanics
literature 
\cite{Dingemans,Peregrine,Jonsson,Chawla,Suastika1,Suastika2},
but to our knowledge
the generation of negative-frequency waves has 
never been observed before.

\section{Negative frequencies}

\begin{figure}[h]
\includegraphics[width=11.5cm]{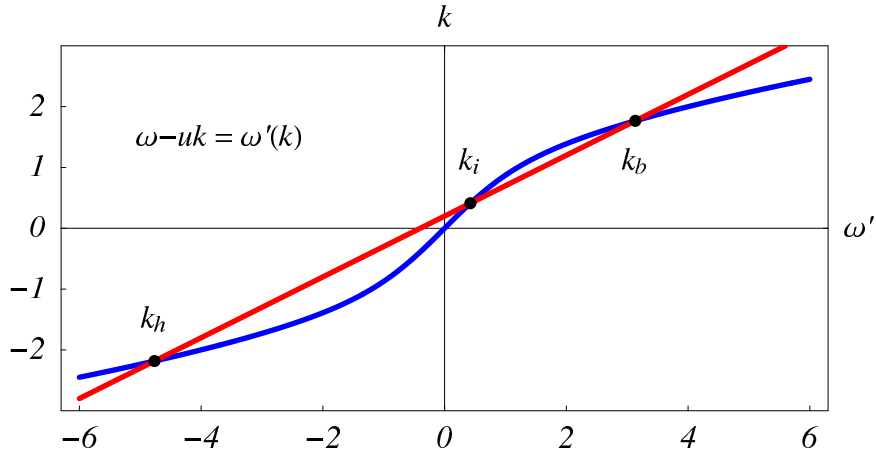}
\caption{{\small
Doppler formula (\ref{eq:doppler}) 
versus dispersion relation (\ref{eq:dispersion}) for $\omega'$
plotted in arbitrary units.
The wavenumber $k_i$ describes the incident wave, 
$k_b$ the blue-shifted and $k_h$ the Hawking wave 
with negative wavenumber $k$ and 
negative frequency $\omega'$.
}
\label{fig:graphical}}
\end{figure}

What are negative-frequency waves?
Consider linear one-dimensional\footnote{The essential physics
of horizons is contained in one-dimensional wave propagation,
even in the case of the three-dimensional black hole,
because near horizons the wavelength is dramatically reduced 
such that their curvatures are insignificant.
}
wave propagation in a moving medium: 
a wave with phase $\varphi$ propagates in the 
$x$ direction against the 
flow $u$. The phase evolves in time $t$ as
\begin{equation}
\varphi = \int \left(k\,\mathrm{d}x-\omega\,\mathrm{d}t\right)    
\label{eq:phase}                                                                
\end{equation}
where $k$ denotes the wavenumber and $\omega$ the frequency
in the laboratory frame.
Imagine we
construct at each point $x$ a frame that is co-moving 
with the fluid.
In the locally co-moving frames\footnote{For simplicity
we ignore effects
of relativistic velocities.}
$\mathrm{d}x=\mathrm{d}x'+ u\,\mathrm{d}t'$
and $\mathrm{d}t=\mathrm{d}t'$,
and so the phase evolves in terms of the co-moving coordinates as
the integral of 
$k\,\mathrm{d}x'-\omega'\,\mathrm{d}t'$                                                              
with
\begin{equation}
\omega' = \omega - uk \,.
\label{eq:doppler}                                                                 
\end{equation}
Equation (\ref{eq:doppler}) simply describes the Doppler effect ---
waves are frequency-shifted due to the motion of the medium.
In a locally co-moving frame, $\omega'$ can only depend
on the wavenumber $k$ and the properties of the medium,
but not explicitly on the position: 
$\omega'$ is a function $\omega'(k)$
that is given by the dispersion relation.
The phase velocity $c'$ is defined as $\omega'/k$,
whereas the group velocity is
\begin{equation}
v_g= \frac{\partial\omega}{\partial k} = v_g'+u \,, \quad
v_g'= \frac{\partial\omega'}{\partial k} \,.
\label{eq:group}                                                 
\end{equation}
What can we say about the dispersion relation in general?
In isotropic media, $\omega'^2$ is an even function of $k$,
because waves should be able to 
propagate in positive and negative directions in the same way.
Without loss of generality we assume  
that the medium moves in the negative direction 
(from the right to the left).
In this case, counter-propagating waves 
have positive phase velocities $c'$. 
Therefore we take the branch of $\omega'$ where 
$\omega'/k$ is positive, {\it i.e.}
where $c'$ is an odd function of $k$ that is positive
for positive $k$.
We also assume that the counter-propagating waves
move with positive group-velocities $v_g'$ in the medium
and that the group-velocity dispersion of the medium 
is normal, {\it i.e.} 
$v_g'$ monotonically decreases for increasing $|k|$.
Figure \ \ref{fig:graphical} shows our specific case
that satisfies these general requirements.

Suppose that the laboratory frequency $\omega$ is fixed.
The wavenumber $k$ is given by
the Doppler formula (\ref{eq:doppler}) and
the dispersion relation $\omega'(k)$.
In general, the solution of this equation is multi-valued:
each frequency $\omega$ corresponds to several wavenumbers $k$,
{\it i.e.} to several physically allowed waves.
As visualized in Fig.\ \ref{fig:graphical},
the physically allowed waves are determined by the points $k$ where
the line $\omega - uk$ intersects the curve $\omega'(k)$. 
One of these wavenumbers $k$ is always negative,
as Fig.\ \ref{fig:graphical} illustrates. 
Since $\omega'$ is an odd function of $k$,
the co-moving frequency $\omega'$ must be negative 
for negative $k$, although the frequency $\omega$
in the laboratory frame is always positive. 
We call waves with negative co-moving frequencies
{\it negative-frequency waves}.
Imagine we display the wave propagation in a space-time diagram,
see Fig.\ \ref{fig:diagrams}.
According to Eq.\ (\ref{eq:phase})
the lines of constant phase $\varphi$ have positive slopes
$\mathrm{d}t/\mathrm{d}x$ for positive $k$
and negative slopes for negative $k$.
We regard this behavior as the characteristic feature 
of negative-frequency waves.

Figure \ref{fig:graphical} shows that
for negative-frequency waves
the slope of the curve $\omega'(k)$ 
is smaller than the slope
of the Doppler line, smaller than $-u$.
As a consequence of Eq.\ (\ref{eq:group}) the group velocity $v_g$
in the laboratory frame must be negative.
Therefore, negative-frequency waves
cannot be launched directly, 
but they can be the result
of a mode conversion from incident positive-frequency waves.

\section{Water waves}

\begin{figure}[h]
\includegraphics[width=12cm]{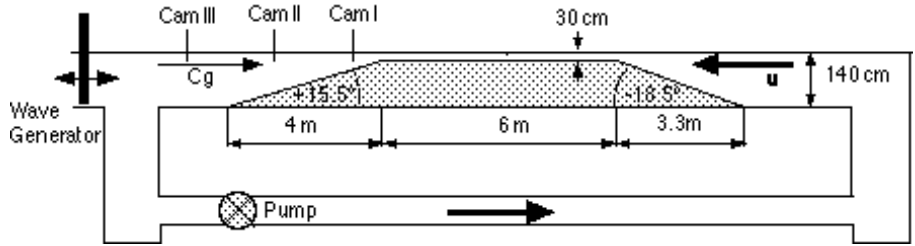}
\vspace*{8pt}
\caption{{\small
Schematic diagram of the experiment.}
\label{fig:scheme}}
\end{figure}

Following a suggestion by Sch\"utzhold and Unruh \cite{SU}, 
we studied water waves in the channel schematically 
shown in Fig.\ \ref{fig:scheme}.
A ramp in the channel creates a gradient in flow speed. 
The flowing water forms a white-hole horizon, 
an object that waves cannot enter, 
when the flow $|u|$ matches the group velocity
$\partial\omega'/\partial k$ of the waves. 
Water waves --- gravity waves --- obey the dispersion relation 
\cite{LL6}
\begin{equation}
\omega'^2 = gk \tanh (kh)
\label{eq:dispersion}
\end{equation}
where $g$ denotes the gravitational acceleration 
of the Earth at the water surface 
and $h$ is the height of the channel. 
In the limit of long wavelengths, 
{\it i.e.} small wavenumbers $k$, 
the dispersion relation (\ref{eq:dispersion}) 
reduces to $\omega'^2 = gh\, k^2$; 
waves propagate with $c' = \sqrt{gh}$. 
We see from the Doppler formula (\ref{eq:doppler}) that, 
in this limit, $\omega'$ is connected to 
$\omega$ and $k$ by a quadratic form, 
which defines a space-time geometry \cite{LL2}. 
A rigorous analysis \cite{SU} proves that the propagation 
of water waves is exactly equivalent to wave propagation 
in space-time geometries, 
as long as $|k|$ is much smaller than $1/h$. 
So, in our case, the channel height $h$ 
serves as a simple analogue of the Planck scale; 
waves with wavelengths shorter than $h$ 
do not experience the effective space-time geometry anymore. 
Close to the horizon, 
the incident waves are compressed 
until $k$ reaches the scale of $1/h$. 

To characterize the waves, 
we use the graphical solution of the Doppler formula 
(\ref{eq:doppler}) 
combined with the dispersion relation (\ref{eq:dispersion}) 
shown in Fig.\ \ref{fig:graphical}. 
For a given positive frequency $\omega$, 
either one or three real solutions exist, 
one negative and possibly two positive $k$. 
Only in the case of a positive solution will the wave-maker 
launch waves, because the group velocity 
(\ref{eq:group}) of the negative-frequency wave is negative.
The slope of $\omega'$ at the smallest positive $k$ 
is higher than the slope of the Doppler line
$\omega - uk$. 
For this wavenumber the group velocity is positive:
this $k$ describes the incident wave. 
When the incident wave propagates against the rising current, 
the slope of the Doppler line rises 
until the two positive $k$ merge. 
At this point, the flow matches the group velocity of the wave. 
The incident wave is converted into a short-wavelength wave; 
it is blue-shifted below the effective Planck scale $h$. 
For the blue-shifted wave, $\partial\omega'/\partial k$ 
lies below the flow speed $|u|$:
the blue-shifted wave 
drifts back with negative group velocity (\ref{eq:group}), 
but $k$ is positive and so is the frequency $\omega'$. 
Figure \ref{fig:graphical} shows that such 
wave blocking 
\cite{Dingemans,Peregrine,Jonsson,Chawla,Suastika1,Suastika2}
cannot occur below a critical flow speed.
In order to estimate \cite{Chawla}
the critical $u$ we replace $\tanh(kh)$
in the dispersion relation (\ref{eq:dispersion})
by the asymptotic value of $1$.
A real $k$ ceases to exist when
the discriminant of the resulting quadratic equation vanishes, 
for $|u|=u^*=g/(4\omega)$.
Since the dispersion curve (\ref{eq:dispersion}) lies below 
the asymptotics,
this procedure \cite{Chawla} gives an overestimation of the critical
flow speed.

The horizon also converts \cite{Unruh2,Brout2,CJ} 
by tunnelling a part of the incident wave into 
the negative-$k$ branch of Fig.\ \ref{fig:graphical} 
that has a positive slope, 
generating a wave with negative co-moving frequency, 
the classical analogue of Hawking radiation. 
In fluid dynamics, the blue-shifted waves 
have been discussed and observed 
in connection with wave-blocking
\cite{Dingemans,Peregrine,Jonsson,Chawla,Suastika1,Suastika2}
but to our knowledge the negative-frequency waves 
have neither been theoretically analyzed 
in the fluid-dynamics literature 
nor experimentally observed.

\section{Experiment}

\begin{figure}[h]
\includegraphics[width=14cm]{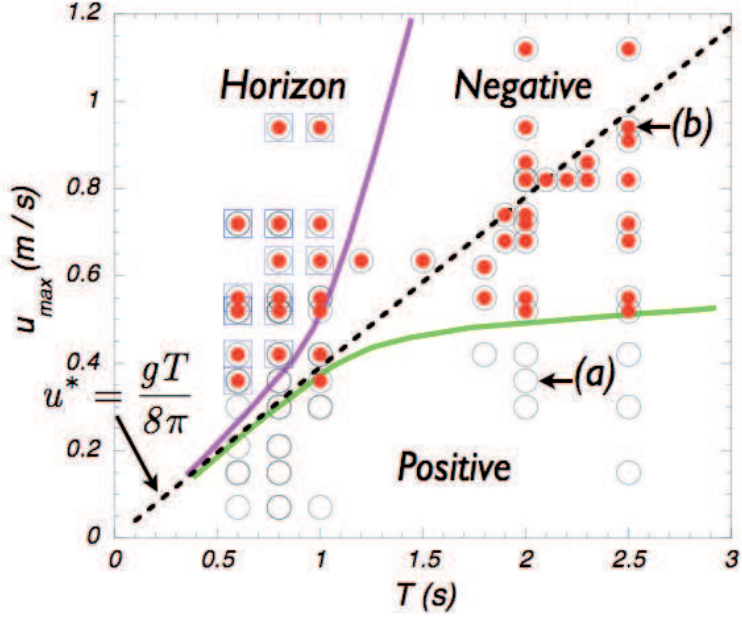}
\caption{{\small 
Phase diagram of our experiment. 
Each circle corresponds to a run with wave period
$T=2\pi/\omega$ and maximal flow speed $u_\mathrm{max}$. 
The dots indicate runs where we observed negative-frequency waves, 
the squares runs with horizons. 
In regimes without horizons we saw a transition 
to mode conversion into purely positive frequencies 
below the lower dotted line in the diagram. 
The points (a) and (b) indicate the parameters used in
Fig.\ \ref{fig:spacetime}.
}
\label{fig:phasediagram}}
\end{figure}

We performed our experiment at ACRI, 
a private research company working on 
environmental fluid mechanics problems such as coastal engineering.
The G\'{e}nimar Laboratory, a department of ACRI, 
features a wave-tank $30\mathrm{m}$ long, $1.8\mathrm{m}$ wide 
and $1.8\mathrm{m}$ deep. 
The wave-maker is of piston-type and can generate waves 
with periods ranging from $0.6\mathrm{s}$ to $2.5\mathrm{s}$ 
with typical amplitudes around $5\mathrm{cm}$ to $30\mathrm{cm}$. 
A current can be superimposed in the same direction 
as the wave propagation or in the opposite one, 
with a maximum flow rate around $1.2\mathrm{m}^3/\mathrm{s}$.
To generate a water-wave horizon, we insert a ramp immersed in water,
with positive and negative slopes separated by a flat section; 
and send on it a train of waves against the reverse fluid flow 
produced by the pump. 
At the place where the flow speed equals the group velocity 
of the waves a horizon is created. 
The geometrical parameters are: 
maximum water height $1.4\mathrm{m}$ or $1.6\mathrm{m}$; 
positive slope $15.5^\circ$; 
length of the flat part $6\mathrm{m}$; 
minimum water height $30\mathrm{cm}$ or $50\mathrm{cm}$; 
negative slope $18.5^\circ$. 
We fix the physical characteristics of the waves, 
period and amplitude, and only vary the background flow. 
We record the waves with the three video cameras indicated 
in Fig.\ \ref{fig:scheme}.
As the background velocity is turbulent 
(the Reynolds number based on the water height is very large) 
and varies with depth, the horizon should be deduced 
from the mean velocity 
$\langle u(h,t) \rangle$
measured at the interface between air and water; 
the brackets denote time averaging. 
Due to experimental constraints, 
we measured the background flow with a MHD sensor 
averaged during $10\mathrm{s}$. 
The velocity profile on the flat part of the background flow 
is plug-like. 
Our first control parameter is $u_\mathrm{max}$, 
the maximum of the counter-current plug velocity 
over the flat part of the geometric profile without water waves. 
We have checked that the velocity profiles are similar along 
a cross section of the tank. 
The second control parameter is the period of oscillations of the wave-maker. 
Both parameters are displayed in the phase diagram of
Fig.\ \ref{fig:phasediagram}.

\begin{figure}[h]
\includegraphics[width=15cm]{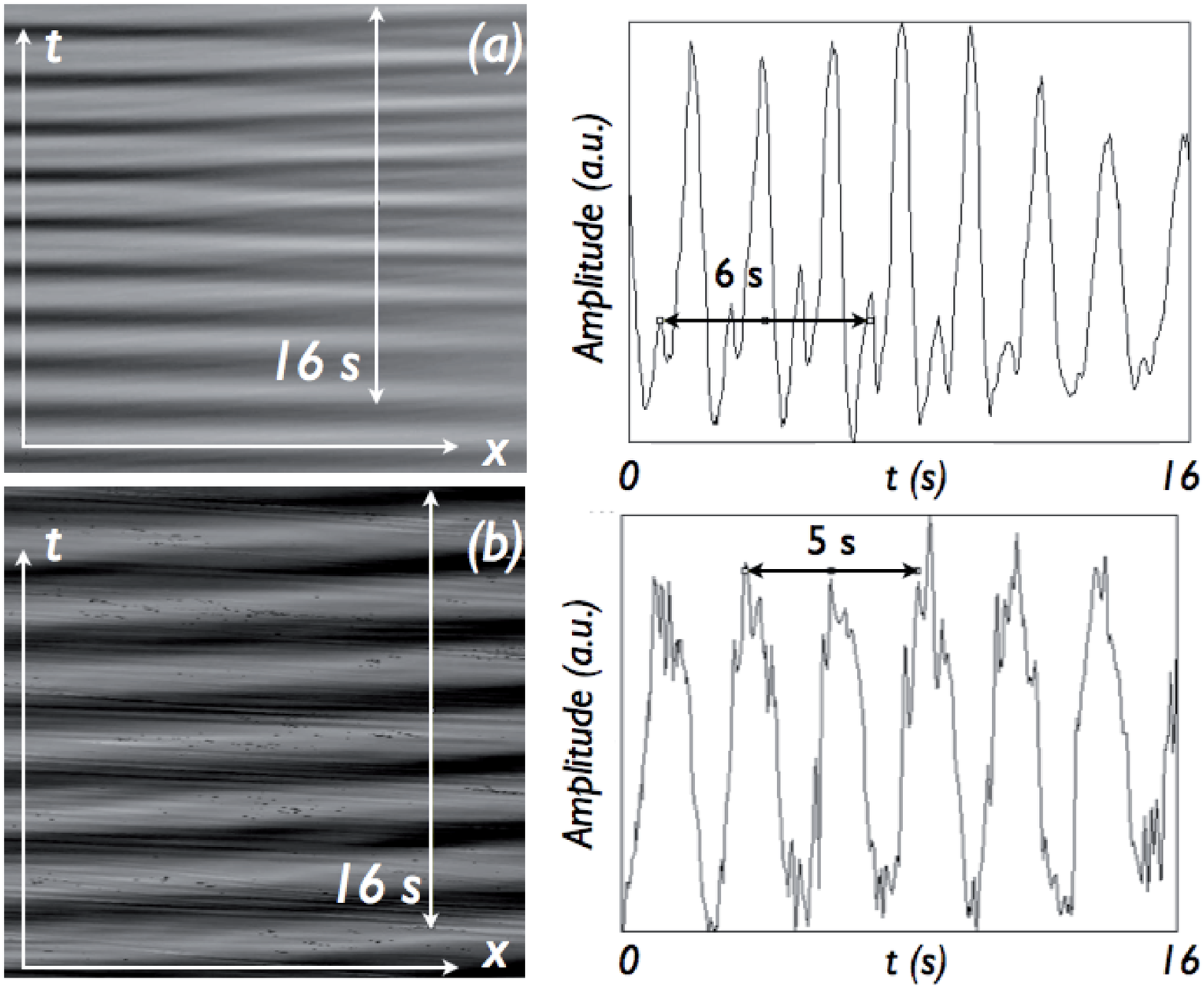}
\vspace*{8pt}
\caption{\small{ 
Space-time diagrams, 
showing water waves propagating 
from the left to the right with the parameters 
(a) and (b) of Fig.\ \ref{fig:phasediagram}, 
initial amplitude $5\mathrm{cm}$ and water height 
$1.4\mathrm{m}$. 
No horizon is formed, but mode conversion still occurs. 
(a) conversion into the positive-frequency waves $k_b$ 
of Fig.\ \ref{fig:graphical}; 
(b) waves with negative frequency 
(negative phase slope as shown in Fig.\ \ref{fig:diagrams}). 
The images were extracted from the video data recorded with camera 1
of Fig.\ \ref{fig:scheme}. 
The right pictures display time traces along the lines 
indicated in the space-time diagrams. 
The traces show that the additional waves are periodic in $T$, 
indicating that they are converted incident waves.}
\label{fig:spacetime}}
\end{figure}

In our experiments, we observed indications of wave conversion 
in the presence of horizons, but the cleanest data 
we obtained was for flow speeds just below the horizon condition. 
In this case, the wave conversion still occurs \cite{Barcelo}, 
although it is reduced in magnitude.
Without a group-velocity horizon, the flow is much quieter, 
wave breaking and turbulence are significantly reduced. 
Figure \ref{fig:spacetime} 
shows the space-time diagrams of two typical cases, 
one illustrating the conversion into short waves 
with positive phase velocity, 
and the other showing waves with negative frequency
superposed on the incident waves. 

\section{Numerical simulations}

\begin{figure}[h]
\includegraphics[width=12cm]{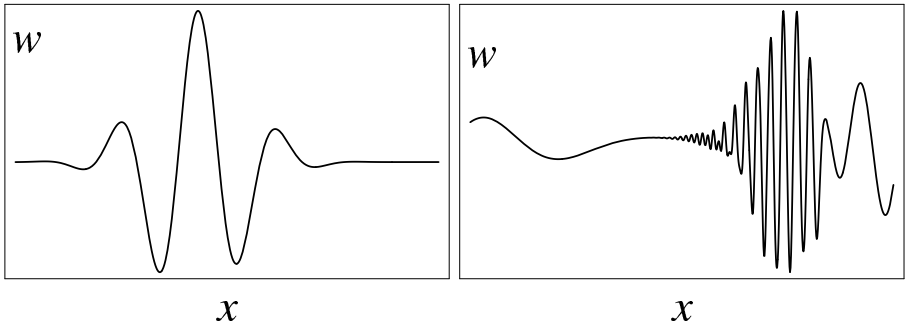}
\vspace*{8pt}
\caption{{\small
Wave-packet simulations. 
The left figure shows the incident wave packet 
traveling in positive $x$ direction, 
the right figure its partial conversion into two wavelength components travelling in negative $x$ direction. 
The components separate because of their different group velocities; the Hawking component is visible in the centre of the figure. 
The wrap-around is caused by periodic boundary conditions 
and most of the packet that travels to the right 
beyond the conversion region is not shown.
We used the parameters 
$u_0 = 0.7\mathrm{m}/\mathrm{s}$, 
$u_1=0.122\mathrm{m}/\mathrm{s}$, 
$a=12\mathrm{m}^{-1}$, 
$h=0.6\mathrm{m}$, 
$T=2.5\mathrm{s}$.}
\label{fig:sim}}
\end{figure}

In order to test whether conversion into
negative-frequency modes occurs 
even in the absence of a horizon,
we applied Unruh's method \cite{Unruh2} 
for numerically simulating waves in moving media.
We consider wave packets propagating against the current 
in a simple one-dimensional model for the flow, 
using periodic boundary conditions, 
and analyse the mode conversion. 
This simulation does not describe the influence of turbulence,
nonlinearity,
the three-dimensional aspects of our experiment 
nor the variation of the flow with water depth, 
but it captures the qualitative aspects of the Hawking effect 
and proves that the mode conversion can occur without a horizon, 
a regime where the experiment 
is least affected by wave breaking and turbulence. 
A related example of Hawking radiation without horizon 
has been studied before \cite{Barcelo} 
that qualitatively agrees with our findings, 
although our case is significantly more extreme. 
Figure \ref{fig:sim} shows the result of a wave packet interacting with the spatially dependent flow given by
\begin{equation}
u(x) = -u_0 - u_1[\tanh(ax) - \tanh(a(x-x_0))] \,;
\end{equation}
the fluid moves left at velocity $-u_0$ at $x<0$, 
decreasing to $-u_0-u_1$ between $x=0$ and $x=x_0$ 
and returning to $-u_0$ at $x>x_0$. 
Gravity waves with the perturbation $w(t,x)$ 
of the velocity potential obey the equation \cite{SU}
\begin{equation}
(\partial_t + \partial_x u)(\partial_t + u \partial_x) w 
= \mathrm{i}g \partial_x 
\tanh (-\mathrm{i}h\partial_x) w \,,
\end{equation}
giving the dispersion relation (\ref{eq:dispersion}). 
The wave packet propagates to the right; 
the flow speed nowhere reaches a value great enough to 
block the packet and create a white-hole horizon. 
When the packet travels into the faster-flow region 
$x>0$ some of it tunnels into the blue-shifted root 
of the dispersion relation and this part propagates back to the left.
There is also some tunnelling into the negative $k$ root; 
this portion has shorter wavelength than the blue-shifted waves 
and travels more quickly to the left.
The simulation shows that negative-frequency waves
can be generated without the presence of a horizon.
The slope in the simulation is not realistic for our experiment, 
however, otherwise there would be no visible 
$k_h$ in the simulation. 
But in the experiment negative-frequency waves 
were clearly observed.
Apparently, the simple model \cite{SU} we used does not
capture all the complexity of our system. 

\section{Conclusions}

We believe we have made the first 
direct observation of the conversion of 
incident waves with positive frequency into 
negative-frequency waves 
in a moving medium.
In astrophysics, such a mode conversion occurs at the event 
horizon of black holes.
It represents the classical mechanism 
at the heart of Hawking radiation \cite{Hawking}.
However,
we were surprised how strong the 
experimentally observed mode conversion is,
because in numerical simulations of a simple model \cite{SU}
we saw a significantly lower conversion.
This model takes into account the correct dispersion relation
(\ref{eq:dispersion}), but it does not describe
turbulence, nonlinearity, nor the three-dimensional 
nature of our experiment. 
It would be highly desirable to find out exactly
what happens to water waves at horizons. 
Unfortunately, 
with the current set-up we have not sufficient data to 
characterize the actual process of mode conversion
in detail.
It is conceivable that we have seen a new fluid-mechanics 
phenomenon that significantly enhances the Hawking effect. 
Could it be a nonlinear mode conversion,
a nonlinear process generating harmonics with
negative frequencies?
We observed that the incident waves 
become steeper as they propagate against the current. 
Hence, locally, waves can be generated close to the crest,
possibly with additional vorticity creation,
where geometric cusps could develop through nonlinear effects. 
These crests waves are then swept away by the flow.\footnote{We 
are indebted to Viktor Ruban for pointing out this mechanism.}
Moreover, it remains to be checked in future experiments whether 
a transverse curvature of the wave crest could also be responsible 
for the creation of negative-frequency waves.
In any case, 
despite the limitations of our present experiment,
we have found clear evidence for 
negative-frequency waves.
In this way,  
we have demonstrated a key ingredient
of the quantum radiation of black holes using 
a relatively simple classical laboratory analogue,
waves in a water tank.

\section*{Acknowledgments}

We thank Philippe Bardey, 
Jean Bougis, 
Mario Novello, 
Renaud Parentani, 
Viktor Ruban
and 
Matt Visser 
for discussions and encouragement, 
and Aurore de Gouvenain, 
Guillaume Bonnafoux, 
Jean-Francois Dest\'{e} and 
Christian Perez 
for technical support. 
This work was financially supported by the Leverhulme Trust
and the University of St Andrews.

\end{document}